\documentclass[aps,prd,preprint,superscriptaddress,tightenlines]{revtex4}
\usepackage{graphicx}% Include figure files
\usepackage{dcolumn}% Align table columns on decimal point
\usepackage{bm}% bold math

\newcommand{\BR}[1]{{\cal B} (#1)}
\newcommand{\ra}{\ensuremath{\rightarrow}}

\newcommand{\Elep}{\ensuremath{E_{\ell}}}
\newcommand{\Enu}{\ensuremath{E_{\nu}}}
\newcommand{\QSqr}{\ensuremath{q^2}}

\newcommand{\MX}{\ensuremath{M_X}}
\newcommand{\MXSqr}{\ensuremath{M_X^2}}
\newcommand{\CosWl}{\ensuremath{\cos\theta_{W\ell}}}

\newcommand{\Vub}{\ensuremath{|V_{ub}|}}
\newcommand{\Vcb}{\ensuremath{|V_{cb}|}}

\newcommand{\Blnu}[1]{\ensuremath{B \ra {#1} l \nu}}
\newcommand{\BXlnu}{\Blnu{X}}
\newcommand{\BXulnu}{\Blnu{X_u}}
\newcommand{\BXclnu}{\Blnu{X_c}}

\newcommand{\BDlnu}{\Blnu{D}}
\newcommand{\BDSlnu}{\Blnu{D^*}}
\newcommand{\BDSSlnu}{\Blnu{D^{**}}}
\newcommand{\BLclnu}{\Blnu{\Lambda_c X}}
\newcommand{\bsg}{\ensuremath{B \ra X_s \gamma\ }}

\newcommand{\BBbar}{\ensuremath{B \overline B}}
\newcommand{\qbar}{\ensuremath{\overline q}}

\newcommand{\BrBXulnu}{\ensuremath{\BR {\BXulnu}}}

\newcommand{\Lambar}{\ensuremath{\overline\Lambda}}
\newcommand{\lam}[1]{\ensuremath{\lambda_{#1}}}

\def\Journal#1&#2&#3(#4){ {\it #1}{\bf #2}, #3 (#4)}

\def\etal{{\it et al.}}

\newcommand{\mypsf}[3]{
\begin{figure}[hptb]
\begin{center}
  #2
\caption{#3}
\label{#1}
\end{center}
\end{figure}
}

\newcommand{\mytable}[3] {
\begin{table}[hptb]
\caption{#2}
\label{#1}
\medskip
\begin{center}
#3
\end{center}
\end{table}
}
\begin{document}

\preprint{CLEO CONF 02-08}
\preprint{ICHEP02 ABS933}

%-------------------------------------------------------------------
% title
%-------------------------------------------------------------------
\title{Preliminary Results on \Vub\ from Inclusive Semileptonic
        $B$ Decays with Neutrino Reconstruction}
\thanks{Submitted to the 31$^{\rm st}$ International Conference on High Energy
Physics, July 2002, Amsterdam}

%-------- INSERT HERE ------------
% Your author list goes here  REMOVE EVERYTHING to END INSERT and
% replace with your authorlist (ask cleoac).

\author{A.~Bornheim}
\author{E.~Lipeles}
\author{S.~P.~Pappas}
\author{A.~Shapiro}
\author{W.~M.~Sun}
\author{A.~J.~Weinstein}
\affiliation{California Institute of Technology, Pasadena, California 91125}                
\author{R.~Mahapatra}
\affiliation{University of California, Santa Barbara, California 93106}                     
\author{R.~A.~Briere}
\author{G.~P.~Chen}
\author{T.~Ferguson}
\author{G.~Tatishvili}
\author{H.~Vogel}
\affiliation{Carnegie Mellon University, Pittsburgh, Pennsylvania 15213}                    
% ...
\collaboration{CLEO Collaboration}
\noaffiliation

%-------- END INSERT ------------

%please hard code the date when you have a final draft and submit to CLEOAC
%\date{\today}
\date{July 23, 2002}

%-------------------------------------------------------------------
% abstract
%-------------------------------------------------------------------
\begin{abstract} 
     We present an analysis of the composition of inclusive semileptonic $B$ meson decays
     using $9.4\ {\rm fb}^{-1}$ of $e^+e^-$ data taken with the CLEO detector at the $\Upsilon(4S)$ 
     resonance. In addition to measuring the charged lepton kinematics, the neutrino 4-vector is 
     inferred using the hermeticity of the detector. We perform a maximum likelihood fit
     over the full three-dimensional differential decay distribution for
     the fractional contributions from the $B\rightarrow X_cl \nu$ processes with $X_c = D$,
     $D^*$, $D^{**}$, and non-resonant $X_c$, and the process $B\rightarrow X_u l \nu$. From
     the fit results we extract 
     $|V_{ub}|= (4.05 \pm 0.18 \pm 0.58 \pm 0.25 \pm 0.21 \pm 0.56) \times 10^{-3}$
     where the errors are statistical, detector systematics , \BXclnu\ model dependence,
     \BXulnu\ model dependence, and theoretical uncertainty respectively.
\end{abstract}

\maketitle

%-------------------------------------------------------------------
% Introduction
%-------------------------------------------------------------------
\section{Introduction}

The CKM matrix element \Vub\ is the coupling between the bottom and up quarks,
and is therefore a fundamental parameter in the Standard Model. Its value
is relevant to studies of  flavor-changing currents, including the rates
of $B$ and $B_s$ mixing and CP violation in the mixing and decays of $B$
hadrons.

Recent theoretical progress in charmless semileptonic $B$ decays using
Heavy Quark Effective Theory (HQET) and the Operator Product Expansion (OPE)
\cite{ref:general_HQET_OPE,ref:falkqsqr,ref:neubertqsqr,ref:bauercuts}
has made possible new  levels of precision in the measurement of \Vub
\cite{ref:Bslendpoint}.
The calculations relate \Vub\ to the partial branching fraction of \BXulnu\ in a 
restricted region of phase space in which \BXclnu\ does not contribute.
The region of phase space most accessible to experiment is the lepton energy endpoint,
$E_\ell >(m_B^2-m_D^2)/2m_B$.
This region was used in the first measurement of \BXulnu\ establishing a non-zero
value of \Vub\cite{ref:firstbtou}. Unfortunately the OPE calculation is not valid
in the very restricted region of phase space delineated by this lepton energy cut.
A recent CLEO measurement has used a spectral function measured in \bsg\ to obtain 
the normalization factor which relates the measured partial branching fraction in 
the lepton endpoint region to \Vub \cite{ref:Bslendpoint}.
Another method is to replace the lepton energy cut with a cut on the invariant mass
of the lepton pair\cite{ref:falkqsqr,ref:neubertqsqr}, \QSqr, or on the mass 
of the recoiling hadronic system, 
\MX, to exclude \BXclnu\ events. Recent theoretical work suggests that a combination
of cuts on both \QSqr\ and \MX\ yields the smallest uncertainty\cite{ref:bauercuts}.

In this paper we report an analysis of \BXlnu\ decays in which both the charged 
lepton and the neutrino are reconstructed. The neutrino is reconstructed using the
approximate hermeticity of the CLEO II
and CLEO II.V detectors and the well known initial state of the $e^+e^-$ system 
produced by the Cornell Electron Storage Ring (CESR).
The \QSqr\ kinematic variable is then calculated directly, and the
\MX\ kinematic variable can be calculated if the $B$ momentum, $\vec{p}_B$,
is known,
\begin{eqnarray*}
\label{mxeqn}
\MX^2 &=& M_B^2 + \QSqr - 2 E_{beam}(E_{\ell}+E_{\nu}) + 2 |\vec{p}_B| |\vec{p}_{\ell\nu}| \cos\theta_{B\cdot \ell\nu},
\end{eqnarray*}
\noindent
where $\vec{p}_{\ell\nu}$ is the sum of the charged lepton and neutrino momenta and $\theta_{B\cdot \ell\nu }$
is the angle between $\vec{p}_{\ell\nu}$ and the $B$ momentum direction.
Since the $B$ mesons are the daughters of an $\Upsilon(4S)$ produced at rest, the magnitude
of the $B$ momentum is known and small, however its direction is unmeasured. 
The last term in the \MXSqr\ equation depends on the $B$ momentum direction, and
is small, unmeasured, and neglected in this analysis.
From the lepton pair sample, the differential decay rate
is measured as a function of the reconstructed quantities \QSqr, \MX, and \Elep.
The experimental resolution on the neutrino four-momentum is poor, with a
narrow core of approximately 120 MeV and broad tail of over-estimation of the
neutrino energy which extends up to 1.5 GeV. The resulting resolution on \QSqr\ and \MX\ 
is also poor, so it is not possible to 
cleanly isolate \BXulnu \ events with cuts on these variables. Instead, we fit this 
observed distribution to a sum
of models for the different hadronic final states: $D$, $D^*$, $D^{**}$, $X_c$ non-resonant,
and $X_u$. From the fit results we infer the partial branching fraction over a region
of phase space with high \QSqr\ and low \MX. These partial branching fractions are
related to \Vub\ by theoretical calculations whose uncertainty has been estimated
\cite{ref:bauercuts}.

The data were accumulated with two configurations of the CLEO detector \cite{ref:NIM},
CLEO II and CLEO II.V, with an integrated luminosity of $9.4\ {\rm fb}^{-1}$ taken on the 
$\Upsilon(4S)$ resonance and an additional $4.5\ {\rm fb}^{-1}$ taken off resonance, 
just below the \BBbar\ production threshold. Both configurations cover 95\% of 
the 4$\pi$ solid angle with drift chambers and a cesium iodide
calorimeter. In addition there are muon chambers with measurements made at 3, 5, and 7 
hadronic interaction lengths of iron and a time of flight system which augments the particle
ID information from specific ionization ($dE/dx$). In the the CLEO II configuration, there were
three concentric drift chambers filled with a mixture of argon and ethane.
In the CLEO II.V detector, the innermost tracking chamber was replaced with a 
three layer silicon detector and the drift chamber gas was changed to a mixture of 
helium and propane.

\section{Neutrino Reconstruction and Event Selection}

Events are selected to have an electron or muon with momentum
greater than 1 GeV/$c$ and a well reconstructed neutrino. Additional cuts are used
to suppress background events from the  $e^+e^- \ra q\qbar$ continuum.

The leptons are selected to fall within the barrel region of the detector 
($|\cos\theta|<.71$, where $\theta$ is the angle between the lepton momentum 
and the beam axis).
Electrons are selected based on $E/p$, $dE/dx$,
and time of flight information combined using a likelihood technique. Muons are
identified by requiring a penetration of 3 interaction lengths of iron for muons
less than 1.5 GeV/$c$ and 5 interaction lengths for muons greater than 1.5 GeV/$c$. The
absolute lepton identification efficiencies are calculated by embedding leptons 
from radiative QED events into hadronic events. The rate at which pions and kaons
fake leptons is measured by reconstructing $K^0_S \ra \pi^+\pi^-$, $D^0 \ra K^-\pi^+$,
and $\bar{D^0} \ra K^+\pi^-$ without using particle identification and then 
checking daughter particle lepton identification signatures.

% fake probs DM5 1% pi->mu, 2% k->mu,
%            DM3 4% pi->mu, 6% k->mu
%            El  < .05% 

Neutrinos are reconstructed by subtracting all observed track and neutral shower
four-momenta from the four-momentum of the $e^+ e^-$ initial state which is at 
rest in the laboratory, 
\begin{eqnarray*}
p_\nu^\mu = p_{e^+ e^-}^\mu - p_{visible}^\mu.
\end{eqnarray*}
The errors made in this assumption are due to particles lost through inefficiency or
acceptance, fake tracks and showers, and other undetected particles such as
second neutrinos, $K^0_L$'s, or neutrons. Several cuts are made to select
events in which these effects are reduced and the neutrino four-momentum resolution
is correspondingly enhanced. Since extra neutrinos are 
correlated with extra leptons, events with a lepton beyond the signal
lepton in the event are rejected.

The primary source of fake
tracks is from charged particles which do not have sufficient transverse momentum,
$p_t$, to reach the calorimeter and therefore curl multiple times in the tracking chambers.
The portions of the track after the initial outbound section may accidentally be called 
a second track. Criteria have been developed to identify such errors and
make a best estimate of the actual charged particles in the event. Events for
which the total charge of the tracks is not zero are removed, reducing the 
effect of lost or fake tracks. 
Showers in the calorimeter associated with tracks in the drift chamber are removed,
so as not to double count their energy. The association between tracks
and showers has been refined to take into account secondary hadronic showers
which can appear to be showers isolated from tracks. 
Finally a requirement is made that the ratio of the reconstructed neutrino four-momentum
divided by twice the reconstructed neutrino energy be less than 0.35 GeV, 
$\frac{{M_{missing}^2}}{2E_\nu}<.35$ GeV. This quantity is constructed to 
be proportional to the energy of a lost or fake particle. The reconstructed neutrino's
energy is assigned to be the magnitude of the momentum, because the momentum is not
dependent on the particle identification of the tracks and so has a better
resolution than the direct energy measurement.

Continuum events are suppressed by a combination of event shape and orientation
variables which exploit the fact that continuum events tend to be jet-like and
aligned with the beam axis, whereas \BBbar\ events are more spherical and
their orientation is uniformly distributed in the detector. 
The second Fox-Wolfram moment \cite{ref:foxwolfram}, R2, of the energy flow in the event is
required to be less than 0.4. In addition a neural network is used to combine R2, 
the angle between the lepton and the thrust axis, 
the angle between the lepton and the beam axis, and the fraction of the energy
lying in 9 separate cones around the lepton direction, which cover the full the $4\pi$
solid angle. The R2 cut is more than 99\% and 95\% efficient for \BXclnu\ and \BXulnu\ respectively,
while removing 60\% of the continuum events. The neural net cut removes an additional
73\% of the continuum background while keeping 92\% and 94\% of the \BXclnu\ and \BXulnu\ 
respectively.

After all cuts we observe 42333 events from CLEO II and 81368 events from CLEO II.V.
The overall efficiency varies from 1.5\% for \BXclnu\ non-resonant to 4.2\% for \BXulnu.

\section{Fitting for Composition}

The full three dimensional differential decay rate distribution as a function
of the reconstructed quantities \QSqr, \MX, and \Elep\ is fit for the contributions
of the signals and backgrounds.
The signals are \BDlnu, \BDSlnu, \BDSSlnu, \BXclnu\  non-resonant, and \BXulnu.
The backgrounds are events where the $B$ decay candidate is faked by a lepton from the
process $b \ra c \ra X \ell\nu$, a lepton from the $e^+e^- \ra q\qbar$ continuum, or a 
fake lepton. We have the freedom to choose the kinematic variables used in the
fit. We use $ \QSqr / (E_\ell + E_\nu )^2$, \MX , and \CosWl.
The helicity angle of the virtual $W$, \CosWl, is defined to be the angle between
the lepton momentum in the $W$ frame and the $W$ in the lab frame. The fit variables have
been chosen to minimize complexity of the phase space boundaries 
and cover the same kinematic space as \QSqr, \Elep, and \Enu.

We perform a binned maximum-likelihood fit where the probability distribution functions (PDFs)
are constructed from weighted Monte Carlo or data events.
The fit uses electrons and muons simultaneously, with a separate set 
of PDFs for the electrons and muons. The likelihood is implemented to take into
account the PDF statistics using the method described in reference \cite{ref:BarlowBeeston}.
The projections of the reconstructed quantities \QSqr, \MX, and \CosWl\ of the various 
\BXlnu\  modes are shown in Figure \ref{fig:btox}.
Projections of the fit result are shown in Figure \ref{fig:fitprojs}.
Projections restricted to the region of enhanced \BXulnu\  sensitivity are shown in 
Figure \ref{fig:btou}. It should be noted that the correlations between the 
the three variables shown in the projections are fully included in the fit, and 
provide considerable power in distinguishing the contributions of the various components.

\mypsf{fig:btox}
{
  \resizebox{.32\textwidth}{!}{\includegraphics{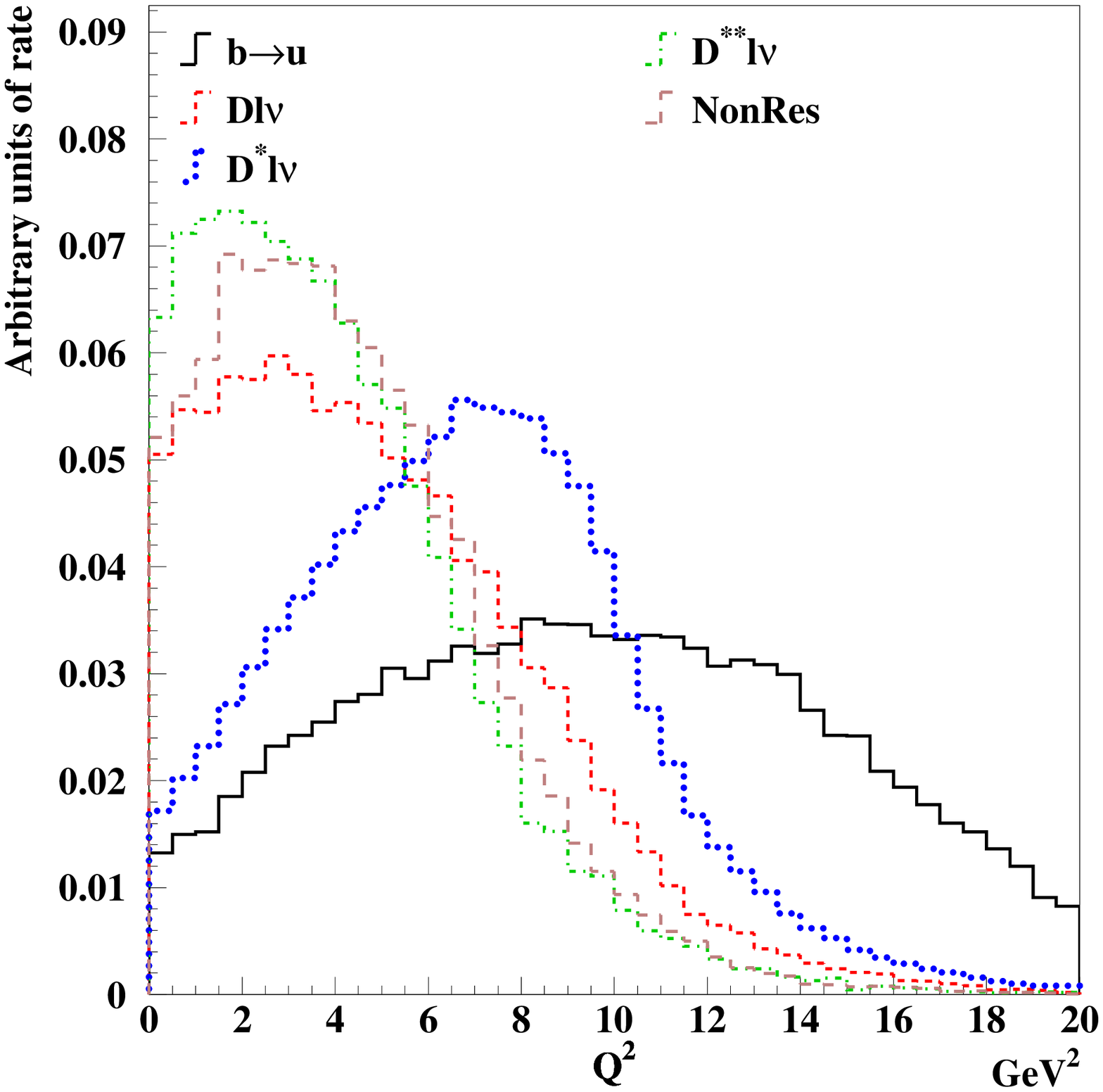}}
  \resizebox{.32\textwidth}{!}{\includegraphics{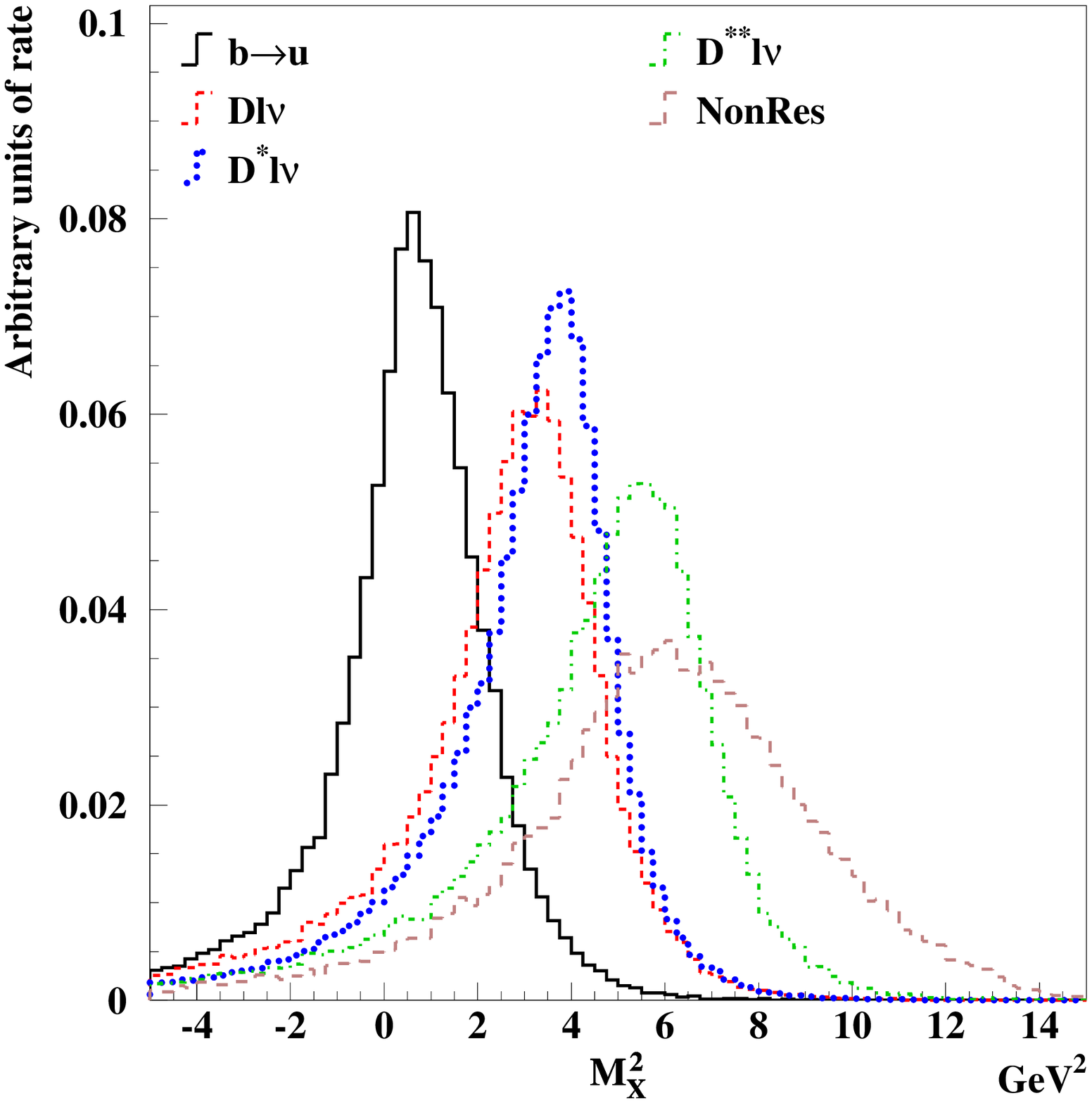}}
  \resizebox{.32\textwidth}{!}{\includegraphics{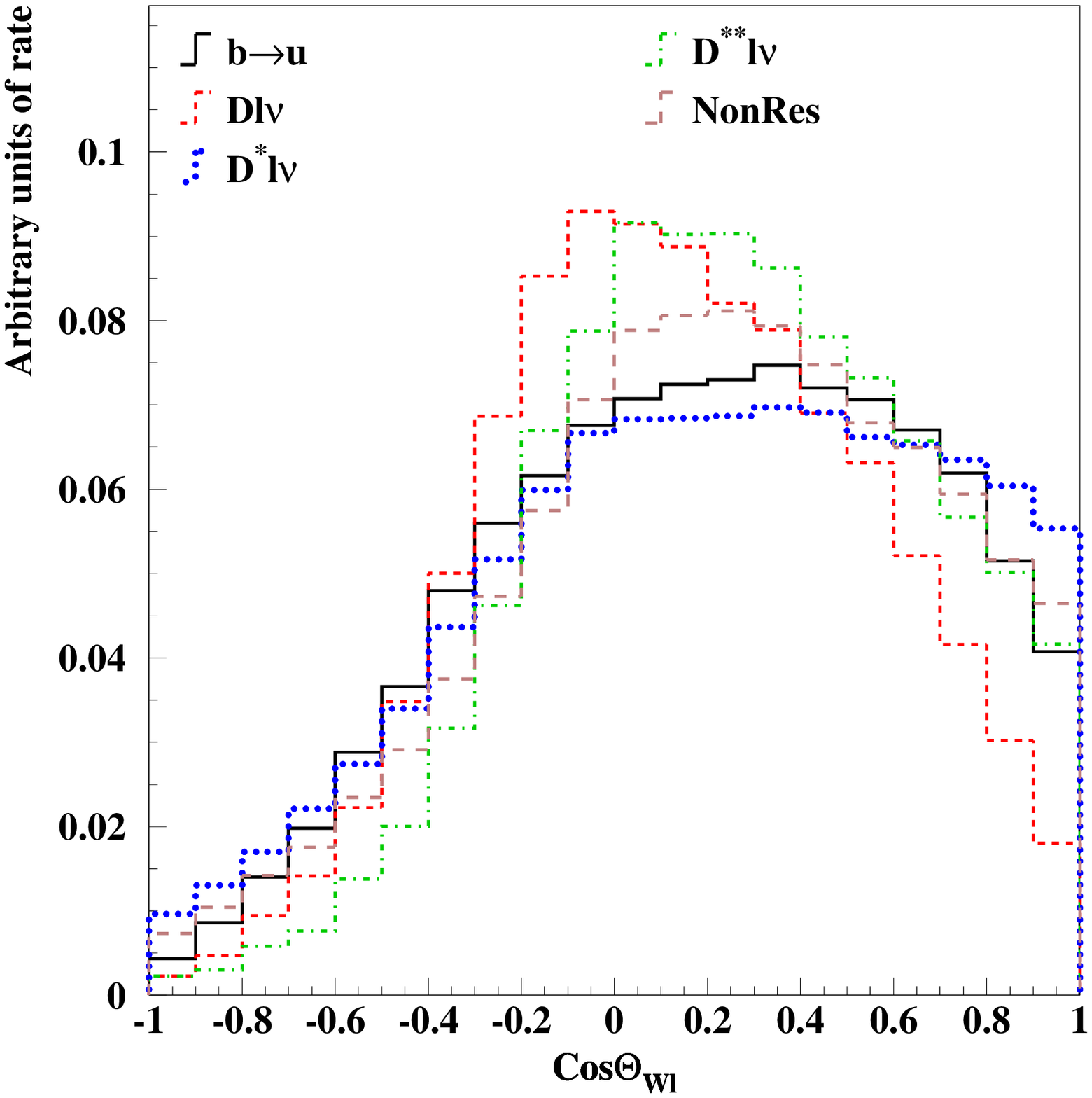}}
}
{
  Distributions of the various \BXlnu\  modes as functions of the reconstructed 
  quantities (left to right) \QSqr, \MXSqr, and \CosWl.
  The modes are \BXulnu\  (solid), \BDlnu\ (short dash), \BDSlnu\ (dots), \BDSSlnu\ (dot-dash),
  and \BXclnu\  non-resonant (long dash).
}

\mypsf{fig:fitprojs}
{
  \resizebox{.32\textwidth}{!}{\includegraphics{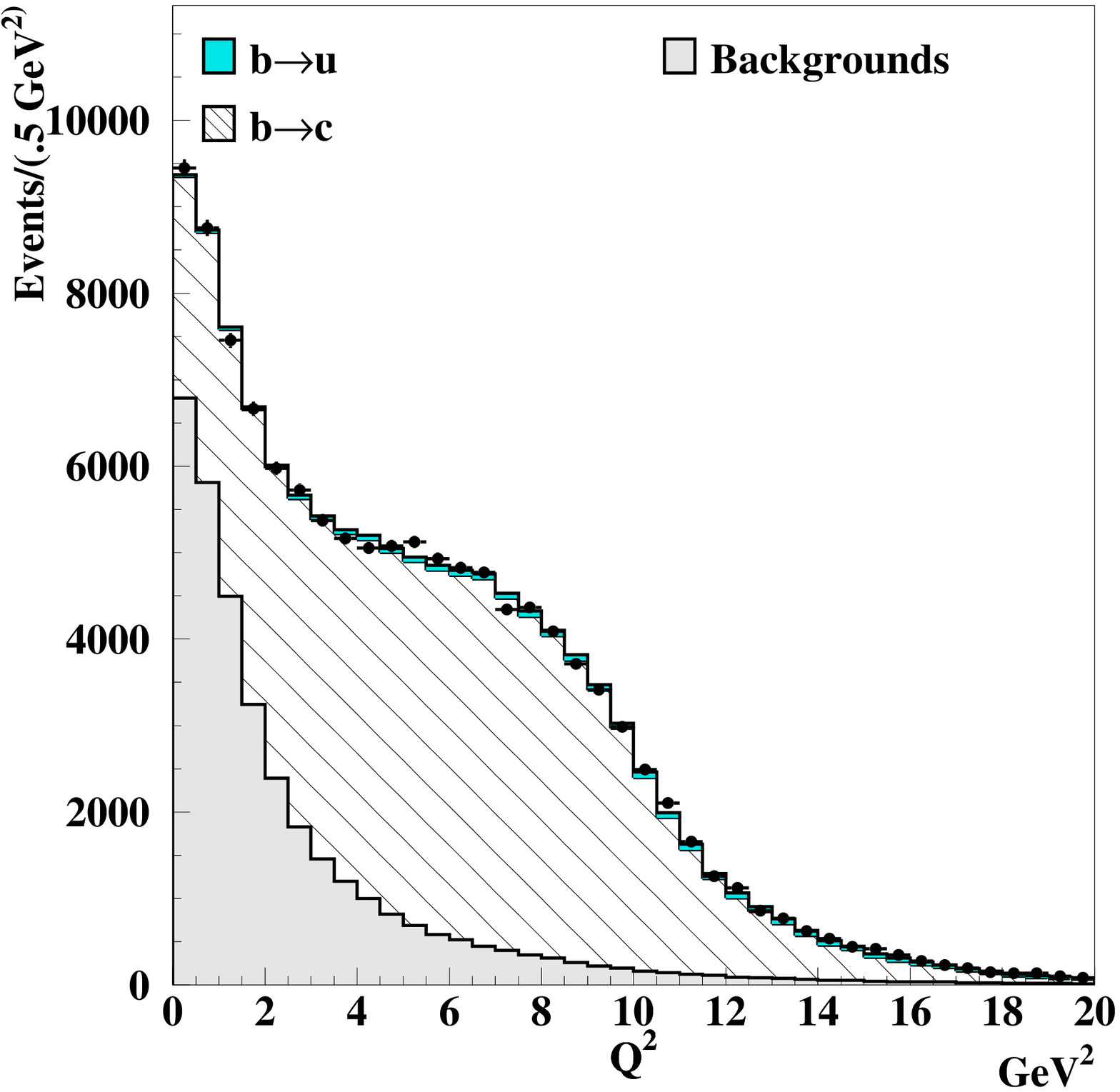}}
  \resizebox{.32\textwidth}{!}{\includegraphics{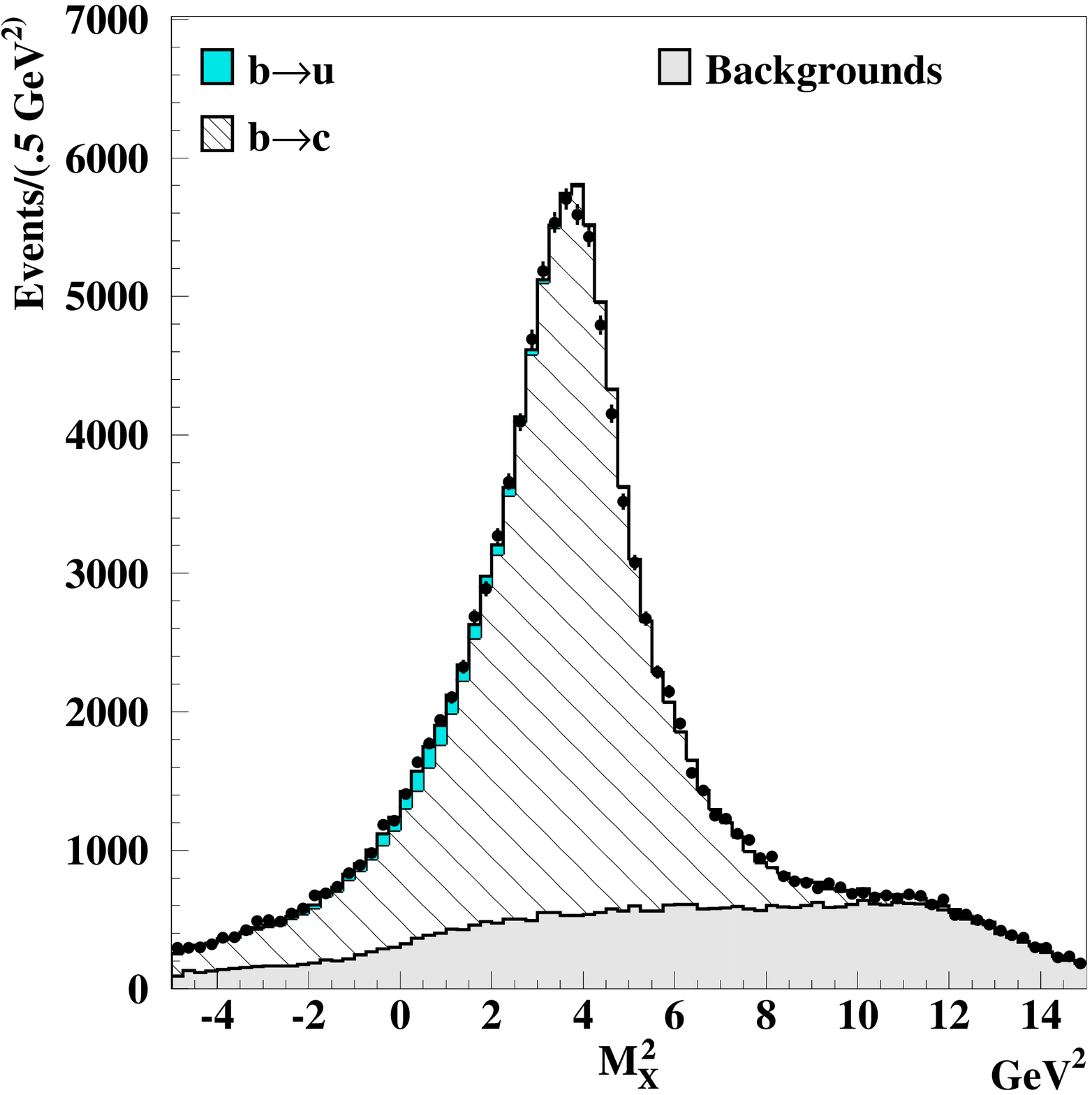}}
  \resizebox{.32\textwidth}{!}{\includegraphics{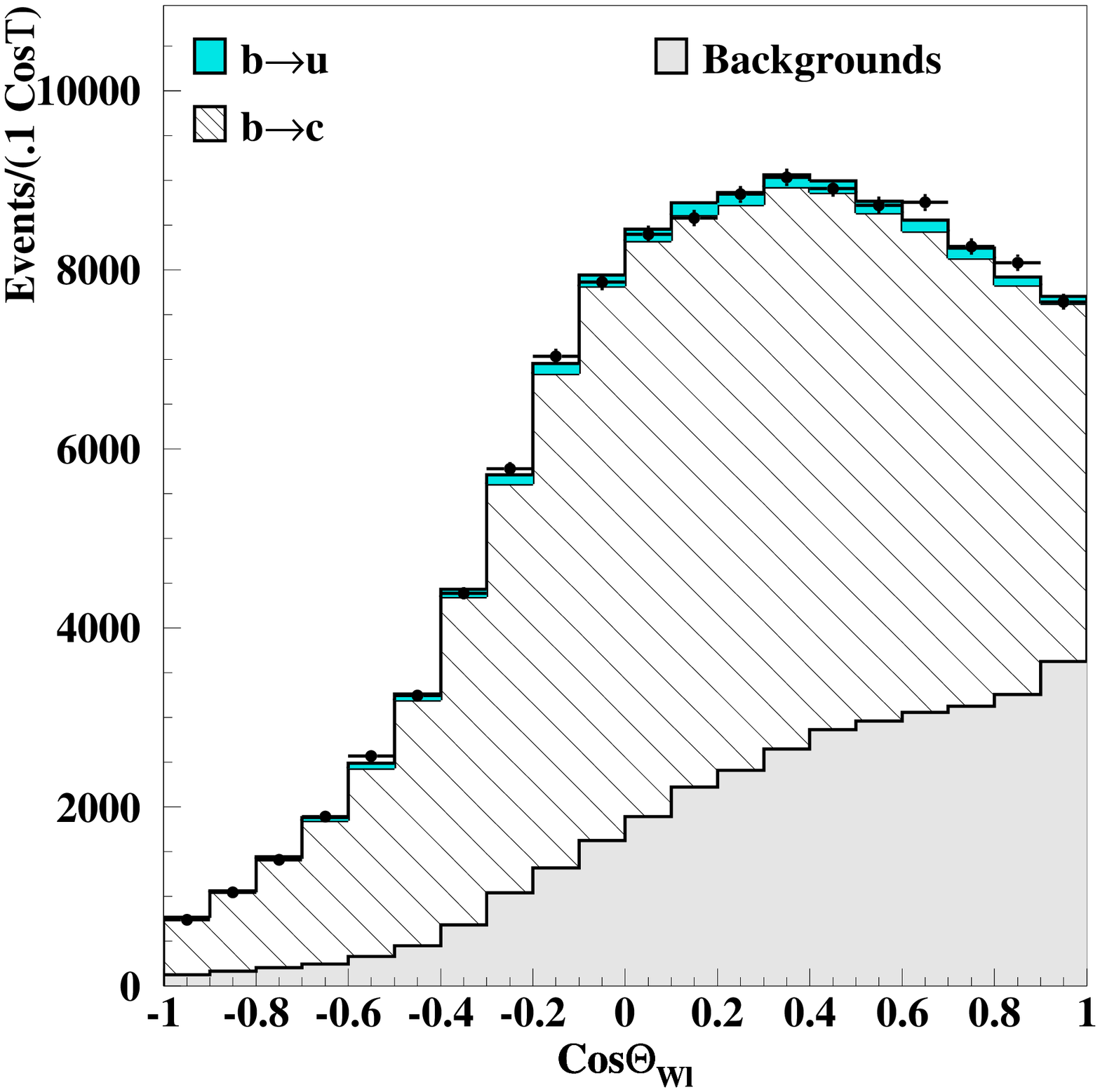}}
}
{
  Projections of the fit results in in the reconstructed quantities  (left to right) \QSqr, \MXSqr, 
  and \CosWl .  The solid histogram on the bottom 
  is the sum of the backgrounds: continuum, secondary leptons, fake leptons. The hatched histogram 
  is the sum of the \BXclnu\  modes: \BDlnu, \BDSlnu, \BDSSlnu, and \BXclnu\  non-resonant. The top
  histogram, barely visible, is the \BXulnu\  component.
}

\mypsf{fig:btou}
{
  \resizebox{.32\textwidth}{!}{\includegraphics{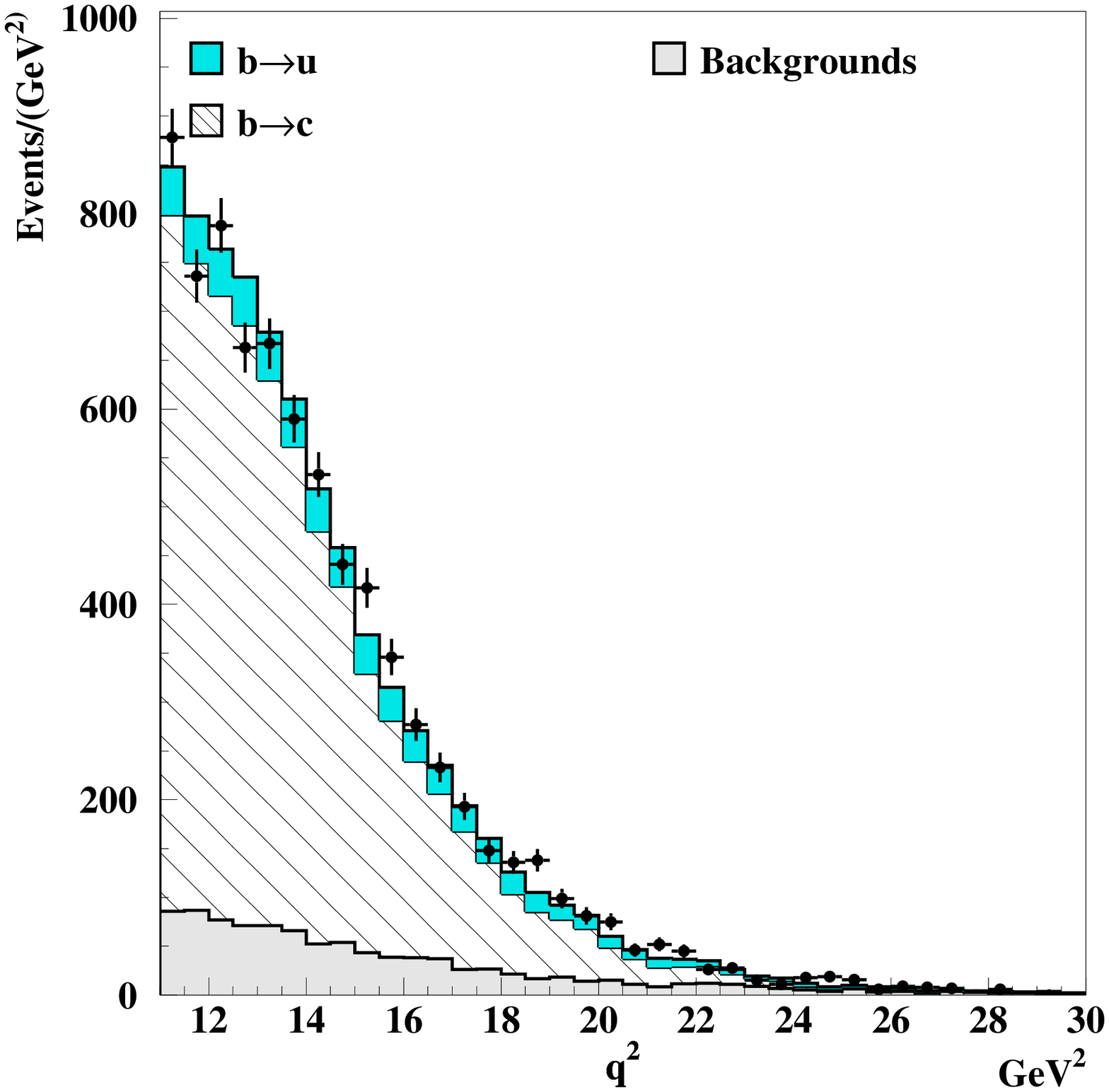}}
  \resizebox{.32\textwidth}{!}{\includegraphics{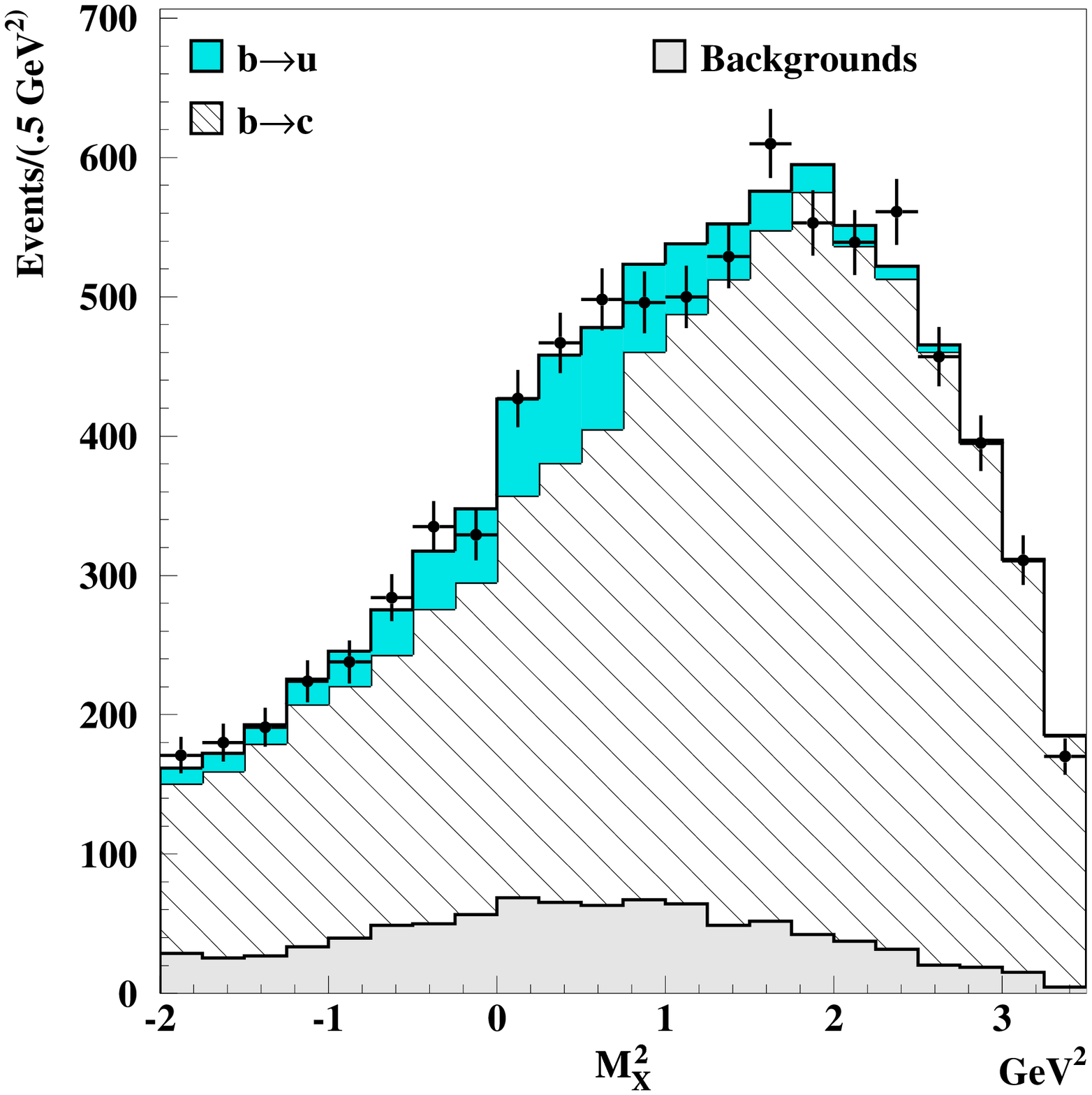}}
  \resizebox{.32\textwidth}{!}{\includegraphics{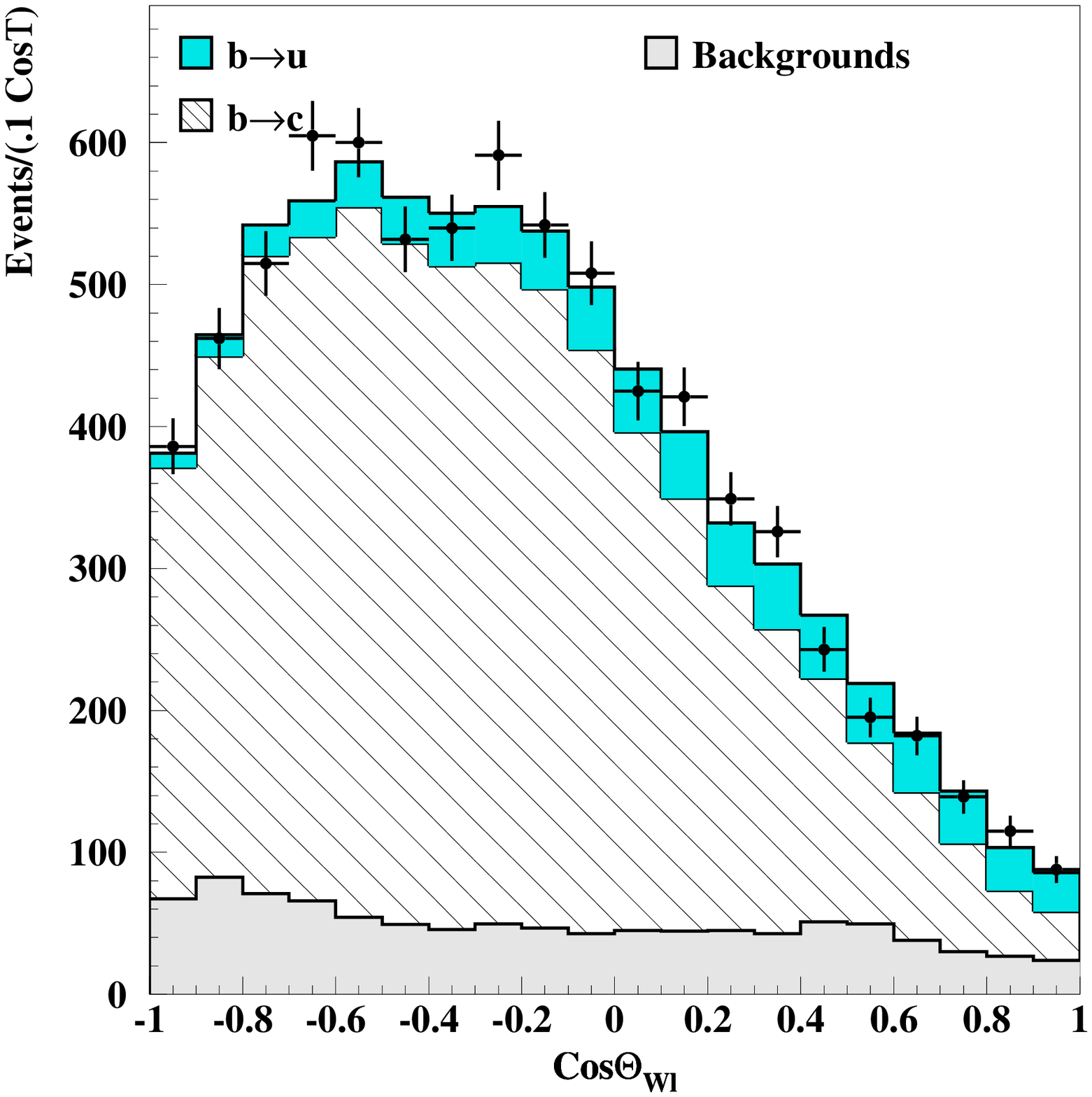}}
}
{
  Projections of the fit results in  reconstructed quantities  \QSqr, \MXSqr, and \CosWl\ 
  with cuts of $\QSqr>11.0\ {\rm GeV}^2/c^4$ for the \MXSqr\ projection and $\MXSqr<2.25\ {\rm GeV}^2/c^4$ for the \QSqr\ 
. projection. The solid histogram on the bottom 
  is the sum of the backgrounds: continuum, secondary leptons, fake leptons. The hatched histogram 
  is the sum of the \BXclnu\  modes: \BDlnu, \BDSlnu, \BDSSlnu, and \BXclnu\  non-resonant. The top
  histogram, now clearly visible, is the \BXulnu\  component.
}

The \BXlnu\ 
modes, secondary leptons and real leptons from continuum are simulated with CLEO's GEANT
based Monte Carlo. The \BDlnu\ and \BDSSlnu\ modes are modeled with ISGW2 \cite{ref:isgw2}.
 The \BDSlnu\ mode is 
modeled with HQET using CLEO's previous measurement of the form factors \cite{ref:DSlnuFFs}.
The $X_c$ non-resonant modes are modeled with a combination of the Goity and Roberts 
model\cite{ref:GoityRoberts} and a crude model of \BLclnu. For an assessment of the \BXclnu\ model
dependence, the \BDlnu\ and \BDSlnu\ form factors are varied within their experimentally 
allowed ranges, and the \BDSSlnu\ masses are varied in an ad-hoc manner. 
Because very little is known about the $X_c$ non-resonant mass spectrum, it is reweighted 
drastically in the study of the \BXclnu\ model dependence. In a further study, each
$X_c$ final state is constrained to a value away from the nominal fit result. The results
of the variations are summed in quadrature. The modeling of the \BXulnu\  mode and 
the resulting model dependence is discussed in a later section.

The fake leptons are modeled by taking a sample of tracks from data that are classified as
$\pi$, $K$, or $\mu$ (electrons and protons are not a significant source of fake leptons)
and unfolding their spectra to get the true spectra of $\pi$, $K$ and $\mu$, which are then
multiplied by the measured fake rates.
This models fake leptons from both \BBbar\ and  $e^+e^- \ra q\qbar$ continuum. The real leptons
from continuum are modeled with continuum Monte Carlo which has been tuned to 
replicate the appropriate charm spectra. Charm is the source of most leptons from 
continuum. The models of both continuum and fakes have been validated and constrained
by a comparison
with the $4.5\ {\rm fb}^{-1}$ of off-resonance data. The secondary leptons are modeled with a 
convolution of CLEO's measured charm spectra from \BBbar\ events and DELCO results on 
the inclusive semileptonic decays of charm\cite{ref:delco}.

The method of neutrino reconstruction adds a large amount of kinematic information
to each event, however it also adds significant potential for systematic errors. The
resolution on the neutrino kinematics is affected by the models of both the signal,
the other $B$ in the event, and the detector response.  The GEANT Monte Carlo does not
perfectly simulate the track and
shower efficiencies and fake rates, nor are $B$ decays well enough understood that the
inclusive particle distributions are well known. For this analysis we employ a method
of reweighting in order to quantify the effects of these uncertainties on our results.
For example to study effect of the tracking efficiency uncertainty on result, the Monte Carlo 
events in which tracks are lost are given a higher or lower weight in constructing the PDFs
and the fit is repeated. The scale of the variation made can in general be constrained 
by direct measurements of the quantity being varied. One important example is that 
having a $K^0_L$ in the event adds a tail to the neutrino resolution. The inclusive $K^0_S$
spectrum in \BBbar\ events has been measured and can be used to constrain the $K^0_L$ spectrum. 

\section{\BXulnu\  Model Dependence}
\label{sec:btoumodel}

We use calculations based on HQET and the OPE to extract \Vub\ from the \BXulnu\ rate
in restricted regions of phase space. This is a well controlled expansion
and the theoretical uncertainties have been assessed by several authors
\cite{ref:falkqsqr,ref:neubertqsqr,ref:bauercuts}. However, the fit's 
region of sensitivity does not coincide with these regions, and, as previously 
mentioned, it is not possible to make cuts to isolate these regions, because 
of the poor resolution on the neutrino four-vector and hence \QSqr\ and \MXSqr.
In order to calculate \Vub, we first make a model dependent inference of the 
partial branching fraction in a region, and then apply the HQET and OPE calculations.
This prescription is designed to minimize the reliance on models and instead
rely on the controlled expansion in HQET and OPE calculations.

\BXulnu\ models are used for two purposes in this analysis. The first 
is in the simulation of the efficiency and resolution of the
detector. This cannot be done with the HQET and OPE calculations,
because they do not predict specific hadronic final states. The 
second use of models is to extrapolate and interpolate between the 
regions of high sensitivity to \BXulnu\ and regions of interest for
extracting \Vub. The fit is primarily sensitive to low \MXSqr\ and 
high \QSqr, but within this region the sensitivity is strongly biased
toward the lepton energy end-point, hence the models are relied on to
extrapolate to the full range of lepton energies.

The result of the fit is the fraction of \BXulnu\ events in the sample. This is converted 
into a branching fraction using the efficiency determined by Monte Carlo simulation
and number of \BBbar s in the sample.
The \BXulnu\ portion of the fit is driven by the region of phase space where the 
\BXclnu\ contribution is not overwhelming, however the fit extrapolates from this region
to the full space using the models. This results in a very large model dependence for
the total rate. The effect can be reduced by calculating for each model a fraction, 
$f_{\rm region}^{\rm model}$, of events in a region of enhanced \BXulnu\ sensitivity. 
The fractions, $f_{\rm region}^{\rm model}$, are calculated from the model using the true
\QSqr\ and \MXSqr, not the reconstructed \QSqr\ and \MXSqr.
The product of the branching fraction from the fit result for a 
model and $f_{\rm region}^{\rm model}$ is the inferred partial branching 
fraction of \BXulnu\ for the model in region selected, 
\begin{eqnarray*}
\Delta {\cal B}_{\rm region}^{\rm model} &=& \BR\BXulnu^{\rm model} \times f_{\rm region}^{\rm model}.
\end{eqnarray*}
$\Delta {\cal B}_{\rm region}^{\rm model}$ has significantly less dependence on the model
than the inferred full branching fraction, $\BR\BXulnu^{\rm model}$, however it still involves
a model dependent combination of interpolation and extrapolation from the actual
region of measurement.

The models used range across extremes from the ISGW2 model 
to a model with only non-resonant modes.
The ISGW2 model contains only low mass hadronic resonances  and no non-resonant or
high mass hadronic systems. The purely non-resonant model implements the HQET and OPE 
prediction for  the hadronic mass spectrum based on the \bsg spectral function,
where the hadronic system is decayed via JETSET. The true spectrum
is likely somewhere between the extremes of all high mass non-resonant and all low
mass resonances. Two permutations of the all non-resonant model  
correspond to changing the spectral function parameters, \Lambar\ and \lam1\, within 
the bounds of the CLEO measurement \cite{ref:Bslendpoint}. Another model combines ISGW2 
and some non-resonant events. In addition, the ISGW2 model is reweighted to have a higher 
or lower \CosWl, corresponding to more or less rate in the lepton end-point, or reweighted 
to have a harder or softer \QSqr\ spectrum.
Table \ref{tab:btoutable} shows the results for the total rate and the inferred partial branching
fractions for five different regions. The central value of $\Delta {\cal B}_{\rm region}$,
reported in the first column of Table \ref{tab:vubtable}, is the center of the range 
of models, and the model dependence uncertainty is assigned to cover the full range.
The fit results for the branching fractions of the final states containing charm 
agree with previous CLEO measurements and will be reported in a subsequent paper.

In the context of HQET and the OPE another fraction, $f_{\rm region}^{\rm HQET}$,
relates the partial branching fraction to \Vub \cite{ref:VubtoTotalRate},
\begin{eqnarray*}
\Vub &=& \left[3.07 \pm 0.12 \times 10^{-3}\right] 
    \left[\frac{\Delta {\cal B}_{\rm region}}{ .001 \times f_{\rm region}^{\rm HQET}}
 \frac{1.6 {\rm ps}}{\tau_B} \right]^{1/2}.
\end{eqnarray*}
The fractions, $f_{\rm region}^{\rm HQET}$, for
the five regions used have been calculated with an evaluation of the theoretical
uncertainty by Bauer {\it et al.} in reference
\cite{ref:bauercuts}. Some of the regions have
also been evaluated in reference\cite{ref:neubertqsqr}.
The \Vub\ results that correspond to the five inferred partial branching fractions
with their respective model dependences are shown in Table \ref{tab:vubtable}.
We choose the $\QSqr>11\ {\rm GeV}^2/c^4,\MX < 1.5\ {\rm GeV}/c^2$ results as a central value because it
has the least model dependence.

\mytable{tab:btoutable}
{
 Dependence on the \BXulnu\  model of the inferred total branching fraction, 
 \BrBXulnu, and the inferred partial branching fractions, 
$\Delta {\cal B}_{\rm region}^{\rm model}$,
 for five regions.  Entries are in units of $10^{-3}$.
}
{
\scriptsize
\begin{tabular}{ccc|ccc|cccc}
      &       & ISGW2 \& &   \multicolumn{3}{|c|}{All Non-Resonant} & &  & Higher & Lower \\
      & ISGW2 & Non-Res & \,Nominal\, & \,Low \Lambar,\lam1\, & \,High \Lambar,\lam1\, & Harder \QSqr & Softer \QSqr & \CosWl & \CosWl \\
\hline
\hline
Total  \BrBXulnu              & 0.996 & 1.211 & 1.808 & 2.289 & 1.620 & 0.803 & 1.159 & 0.956 & 1.039 \\  
$\QSqr>6\ {\rm GeV}^2,\MX < M_D$      & 0.629 & 0.686 & 0.835 & 0.859 & 0.811 & 0.582 & 0.655 & 0.606 & 0.653 \\ 
$\QSqr>8\ {\rm GeV}^2,\MX < 1.7\ {\rm GeV}$   & 0.497 & 0.528 & 0.609 & 0.617 & 0.598 & 0.482 & 0.494 & 0.480 & 0.516 \\ 
$\QSqr>11\ {\rm GeV}^2,\MX < 1.5\ {\rm GeV}$  & 0.308 & 0.317 & 0.348 & 0.347 & 0.346 & 0.321 & 0.283 & 0.297 & 0.319 \\ 
$\QSqr>(M_B-M_{D^{*}})^2$     & 0.283 & 0.293 & 0.329 & 0.338 & 0.318 & 0.298 & 0.257 & 0.273 & 0.293 \\ 
$\QSqr>(M_B-M_D)^2$           & 0.331 & 0.344 & 0.393 & 0.410 & 0.377 & 0.342 & 0.307 & 0.319 & 0.343 \\ 
\hline
\end{tabular}
}
\mytable{tab:vubtable}
{
 Inferred partial branching fraction by region, $\Delta {\cal B}_{\rm region}$, with
 model dependence in the first column. \Vub\ calculated from the inferred partial branching
 fractions with all errors in the second column. The errors on \Vub\ are statistical, detector,
 \BXclnu\  model dependence, \BXulnu\  model dependence, and theoretical uncertainty respectively.
 Please note, these results do {\em not} come from fits in the restricted regions.
}
{
\scriptsize
\begin{tabular}{cc|c}
      & ~ $\Delta {\cal B}_{\rm region} \pm$ $b\ra u$ Model Error  ~
      & ~\Vub $\pm$ Stat $\pm$ Detector $\pm$  $b\ra c$ $\pm$ $b\ra u$ $\pm$ Theory ~\\
\hline
\hline
Total \BrBXulnu              & $(1.546 \pm 0.743) \times 10^{-3}$ & $(3.82 \pm 0.17 \pm 0.55 \pm 0.23 \pm 0.92 \pm 0.12) \times 10^{-3}$ \\
$\QSqr>6\ {\rm GeV}^2,\MX < M_D$     & $(0.720 \pm 0.138) \times 10^{-3}$ & $(3.84 \pm 0.17 \pm 0.55 \pm 0.23 \pm 0.37 \pm 0.31) \times 10^{-3}$ \\
$\QSqr>8\ {\rm GeV}^2,\MX < 1.7\ {\rm GeV}$  & $(0.548 \pm 0.069) \times 10^{-3}$ & $(3.98 \pm 0.18 \pm 0.57 \pm 0.24 \pm 0.25 \pm 0.38) \times 10^{-3}$ \\
$\QSqr>11\ {\rm GeV}^2,\MX < 1.5\ {\rm GeV}$ & $(0.315 \pm 0.032) \times 10^{-3}$ & $(4.05 \pm 0.18 \pm 0.58 \pm 0.25 \pm 0.21 \pm 0.56) \times 10^{-3}$ \\
$\QSqr>(M_B-M_{D^{*}})^2$    & $(0.297 \pm 0.041) \times 10^{-3}$ & $(4.07 \pm 0.18 \pm 0.58 \pm 0.25 \pm 0.28 \pm 0.62) \times 10^{-3}$ \\
$\QSqr>(M_B-M_D)^2$          & $(0.359 \pm 0.052) \times 10^{-3}$ & $(4.05 \pm 0.18 \pm 0.58 \pm 0.25 \pm 0.29 \pm 0.54) \times 10^{-3}$ \\
\hline
\end{tabular}
}

\section{Conclusion}

We have made a preliminary measurement the CKM parameter \Vub\ with high precision,
but some model dependence, and obtain
\begin{eqnarray*}
V_{ub} &=& (4.05 \pm 0.18 \pm 0.58 \pm 0.25 \pm 0.21 \pm 0.56) \times 10^{-3}
\end{eqnarray*}
where the errors are statistical, detector systematics, \BXclnu\ model dependence,
\BXulnu\ model dependence, and theoretical uncertainty respectively.
The result is consistent with the CLEO result using the lepton end-point measurement
\cite{ref:Bslendpoint}. There is some statistical correlation with the lepton end-point 
measurement and the model dependence may be correlated as well. These correlations will 
be investigated. This analysis is the first to use neutrino reconstruction and the full 
three-dimensional kinematic information to extract the \BXulnu\ branching fraction. This 
also pioneers the application of the multidimensional cuts laid out by Bauer \etal\cite{ref:bauercuts}. A more detailed study of the region of sensitivity is in progress, 
and may result in a less model dependent extraction of \Vub.

The composition of the \BXclnu\ is also currently being analyzed and will provide new 
information on the branching fractions, the HQET parameters \Lambar\ and \lam1, and the 
Standard Model parameter \Vcb.

We gratefully acknowledge the contributions of the CESR staff for providing the luminosity
and the National Science Foundation and U.S. Department of Energy for supporting this research.

\end{document}